# Predicting missing values in spatio-temporal satellite data

Florian Gerber[a], Reinhard Furrer[a], Gabriela Schaepman-Strub[b], Rogier de Jong[c], Michael E. Schaepman[c]

[a] Institute of Mathematics, University Zurich, Winterthurerstrasse 190, CH-8057 Zurich, Switzerland, florian.gerber@math.uzh.ch and reinhard.furrer@math.uzh.ch

[b] Department of Evolutionary Biology and Environmental Studies, University of Zurich, Winterthurerstrasse 190, CH-8057 Zurich, Switzerland, gabriela.schaepman@ieu.uzh.ch

[c] Remote Sensing Laboratories, Department of Geography, University of Zurich, Winterthurerstrasse 190, CH-8057 Zurich, Switzerland, michael.schaepman@geo.uzh.ch and rogier.dejong@geo.uzh.ch

Corresponding author:

Reinhard Furrer, Institute of Mathematics, University Zurich, Winterthurerstrasse 190, CH-8057 Zurich, Switzerland, reinhard.furrer@math.uzh.ch

**Highlights:**

- Flexible and scalable gap-fill algorithm for remotely sensed data
- Tested with MODIS NDVI data featuring up to 50% missing values
- Validated against established software
- Uncertainty quantification of the predicted values
- Software and examples provided in the open-source R package *gapfill*

# Abstract

Remotely sensed data are sparse, which means that data have missing values, for instance due to cloud cover. This is problematic for applications and signal processing algorithms that require complete data sets. To address the sparse data issue, we present a new gap-fill algorithm. The proposed method predicts each missing value separately based on data points in a spatio-temporal neighborhood around the missing data point. The computational workload can be distributed among several computers, making the method suitable for large datasets. The prediction of the missing values and the estimation of the corresponding prediction uncertainties are based on sorting procedures and quantile regression. The algorithm was applied to MODIS NDVI data from Alaska and tested with realistic cloud cover scenarios featuring up to 50% missing data. Validation against established software showed that the proposed method has a good performance in terms of the root mean squared prediction error. The procedure is implemented and available in the open-source R package *gapfill*. We demonstrate the software performance with a real data example and show how it can be tailored to specific data. Due to the flexible software design, users can control and redesign major parts of the procedure with little effort. This makes it an interesting tool for gap-filling satellite data and for the future development of gap-fill procedures.

# Graphical abstract

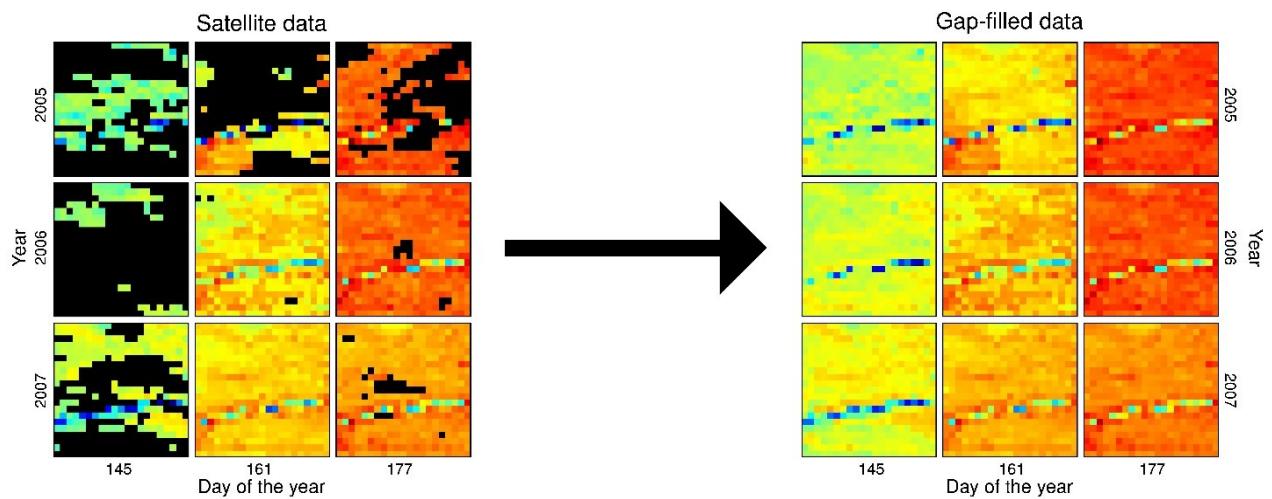

# 1. Introduction

Remote sensing is a technology used to study a wide range of Earth surface processes. It is usually used a long distance from the ground, and when compared to ground based measurements, the technology has the advantage of large spatial and temporal coverage. It is, however, crucial to understand and correct occasional measurement errors, which are caused by off-nadir view angles and atmospheric disturbances. We take a closer look at the data workflow of satellite observations to understand how they influence the study of satellite observations. To correct for any inaccuracies or omissions, we introduce a new gap-filling algorithm that assuages any discrepancies.

## 1.1. Missing values in satellite data

One typical example of a remote-sensing data workflow that deals with missing values is the analysis of the Earth's vegetation using measurements from the Moderate Resolution Imaging Spectroradiometer (MODIS) sensor on-board the Terra and Aqua platforms. The sensors measure radiation reflected by the Earth surface every one to two days. Atmospheric disturbances like cloud cover and sub-optimal view angles introduce observation error; data, in these cases, may have measurement error or missing data for parts of the scene (e.g. Myneni et al., 2002). To retrieve the desired information, the data are pre-processed and transformed into a data product; for example the MOD13 family of vegetation indices. This pre-processing phase consists of aggregation techniques such as constrained-view-angle maximum-value composites (van Leeuwen, Huete, & Laing, 1999) and quality assignments (Roy et al., 2002). A resulting data product is MOD13A (Didan, Munoz, Solano, & Huete, 2015), which comprises several data layers containing values on a regular grid in space and time with a resolution of 500 m and 16 days. However, when only values classified with "good quality" are considered, the proportion of missing data can be considerable (up to 100% of the pixels for certain images of the Alaska region). This may negatively influence the inference of Earth-surface processes or even render certain analysis techniques infeasible.

The strategies that handle the missing data problem can be divided into two groups. The first group employs statistical data analysis methods that are robust to missing values. The second group predicts the missing values, in the data, using an additional processing step before the analysis. This can either be done for a data product containing missing values or as an integral part of the pre-processing.

## 1.2. Existing approaches to handle missing values

In the following overview, existing methods that handle satellite data with missing values are grouped into robust analysis methods and methods that reconstruct a complete data product before the actual data analysis.

### 1.2.1. Robust analysis methods

When describing vegetation using remotely sensed data, the temporal characterization of the process is of great interest. This includes short and long-term trends of the values themselves and of derived quantities such as growing season onset, growing season length, etc. Many methods exist to study these phenomena by interpreting the satellite data as a collection of spatially-independent time series (one per pixel in the spatial extent). Commonly, the time series are smoothed to reduce noise and to fill the missing values. The information of interest is then extracted from this smoothed time series. Hence, the presence of missing values is no longer a problem, as long as the smoothing is not negatively influenced by the missing values. Examples of smoothing methods are the Savitzky-Golay filter method (Chen et al., 2004), the least-squares fitted asymmetric Gaussian method and the double logistic smooth functions method, which are all implemented in the software TIMESAT (Jönsson & Eklundh, 2004). Other approaches use Fourier analysis (Dash, Jeganathan, & Atkinson, 2010; Roerink, Menenti, & Verhoef, 2000; Scharlemann et al., 2008), locally weighted scatterplot smoothing (LOESS) (Moreno, Garcia-Haro, Martinez, & Gilabert, 2014) and the CACAO method (Verger, Baret, Weiss, Kandasamy, & Vermote, 2013). The software TiSeG (Colditz, Conrad, Wehrmann, Schmidt, & Dech, 2008) retrieves vegetation time-series though interpolation in combination with pixel-level quality assurance data. Many studies compared these and similar methods by applying them to simulated and observed data (Atkinson, Jeganathan, Dash, & Atzberger, 2012; Atzberger & Eilers, 2011; Gao et al., 2008; Hird & McDermid, 2009; Kandasamy, Baret, Verger, Neveux, & Weiss, 2013; Musial, Verstraete, & Gobron, 2011; Neteler, 2010). Depending on the test scenarios and the comparison techniques, different methods performed better.

Moreover, specialized methods exist to study local trends in vegetation index time series, e.g., the "breaks for additive season and trend" (BFAST) algorithm (Verbesselt, Hyndman, Newnham, & Culvenor, 2010; Verbesselt, Hyndman, Zeileis, & Culvenor, 2010) and the "detecting breakpoints and estimating segments in trend" (DBEST) algorithm (Jamali, Jönsson, Eklundh, Ardö, & Seaquist, 2015). While all of these methods are based on the temporal correlation of the values, Bolin et al. (2009) presented a method that exploits both the temporal and spatial correlation to study spatial patterns of temporal trends in NDVI data.

### 1.2.2. Reconstructing a complete data product

Another strategy to handle missing values is to predict them based on the observed data. This is typically done in an additional step before the analysis of interest. Several authors discuss the use of geostatistical methods such as Kriging and Co-Kriging for this task (Addink, 1999; Rossi, Dungan, & Beck, 1994). These methods exploit the spatial correlation of the data within an image. In the case of Co-Kriging information from images observed at different points in time are included as a regression terms. The same ideas were used to restore Landsat ETM+ data that exhibit systematically missing strips caused by the SLC failure in 2003 (Pringle, Schmidt, & Muir, 2009; C Zhang, Li, & Travis, 2007; Chuanrong Zhang, Li, & Travis, 2009). The Kriging ideas were extended in the direction of spatio-temporal models via the combination with generalized additive models (Poggio, Gimona, & Brown, 2012) and Kalman-filtering (Lguensat, Tandeo, Fablet, & Garello, 2014).

In contrast to the aforementioned methods, the following approaches are not derived from the classical geostatistical framework. Chen at al. (2011) proposed a method known as "neighborhood similar pixel interpolator", which uses weighted values from images observed at other points in time. This approach was applied to Landsat ETM+ SLC failure data (Mohammdy et al., 2014) and combined with a Kriging (Zhu, Liu, & Chen, 2012). Based on similar ideas, an algorithmic gap-fill procedure for MODIS EVI data was derived and tested at continental scale (Weiss et al., 2014). Other methods use linear regression models fitted to a spatio-temporal window around the missing pixel (J C de Oliveira & Epiphanio, 2012; Julio Cesar de Oliveira, Epiphanio, & Renno, 2014). These methods rely on additional land cover data (Kang, Running, Zhao, Kimball, & Glassy, 2005; Maxwell, Schmidt, & Storey, 2007), or use the quality information provided for MODIS data products to predict missing values (J. Gu, Li, Huang, & Okin, 2009; Y. Gu, Bélair, Mahfouf, & Deblonde, 2006).

### 1.3. Outline

We introduce and discuss a new gap-fill procedure. This procedure can be applied to reconstruct a complete data product, before analysis of satellite data. Similar to the methods presented by Weiss et al. (2014), de Oliveira & Epiphanio (2012), and de Oliveira et al. (2014), we use spatio-temporal subsets around the missing values to predict them. Our contributions are that we (1) formalize this subset-predict procedure in a generic way and design the software implementation of the algorithm accordingly, (2) present a new instance of such a subset-predict algorithm, which provides very accurate fill values for our test scenarios, and (3) base the proposed algorithm on a statistical frame-work, which helps to quantify the uncertainties associated with the predicted values.

We introduce the gap-fill method, the used validation methods and test datasets in the following sections. The performance of the proposed procedure is then assessed with four test scenarios and a realistic MODIS NDVI data example. In addition, we compare the accuracy of the predicted values against those obtained with the TIMESAT software and with the "Gapfill-Python" program implementing the algorithm described in Weiss et al. (2014).

## 2. Gap-fill method

### 2.1. Illustration of the concept

The gap-fill method is tailored to data that have the typical four-dimensional array structure of satellite products. This is, two dimensions are used to describe the spatial location of a value, e.g., through longitude and latitude. The other two dimensions describe its temporal position through a seasonal index and the year. For example, if the data are available on a 16-day time interval basis, the seasonal index can take on values 1 to 24 and indicates the time interval of the measurement within the year. We illustrate the prediction algorithm with the MODIS NDIV data (satellite product MOD13A1) shown in panel (a) of Figure 1. Note, however, that the proposed method can be applied to a wide range of remotely sensed data. The example data have a spatial extent of $21 \times 21$ pixels and consists of $16$ images having $4$ seasonal indices (the days 145, 151, 177, and 193 of the year) observed over $4$ years (2004—2007). With that, the data array has $21 \times 21 \times 16 = 7'056$ elements in total. A characteristic of MOD13 data is that the values exhibit spatial and temporal dependencies. This is usually the case for (pre-processed) satellite data and can also be observed in the example data. We distinguish between observed and missing values. In panel (a) of Figure 1, $1'603$ ($\approx 23\%$) values are missing and depicted as black pixels. The goal of the gap-filling procedure is to predict the missing values from the observed ones.

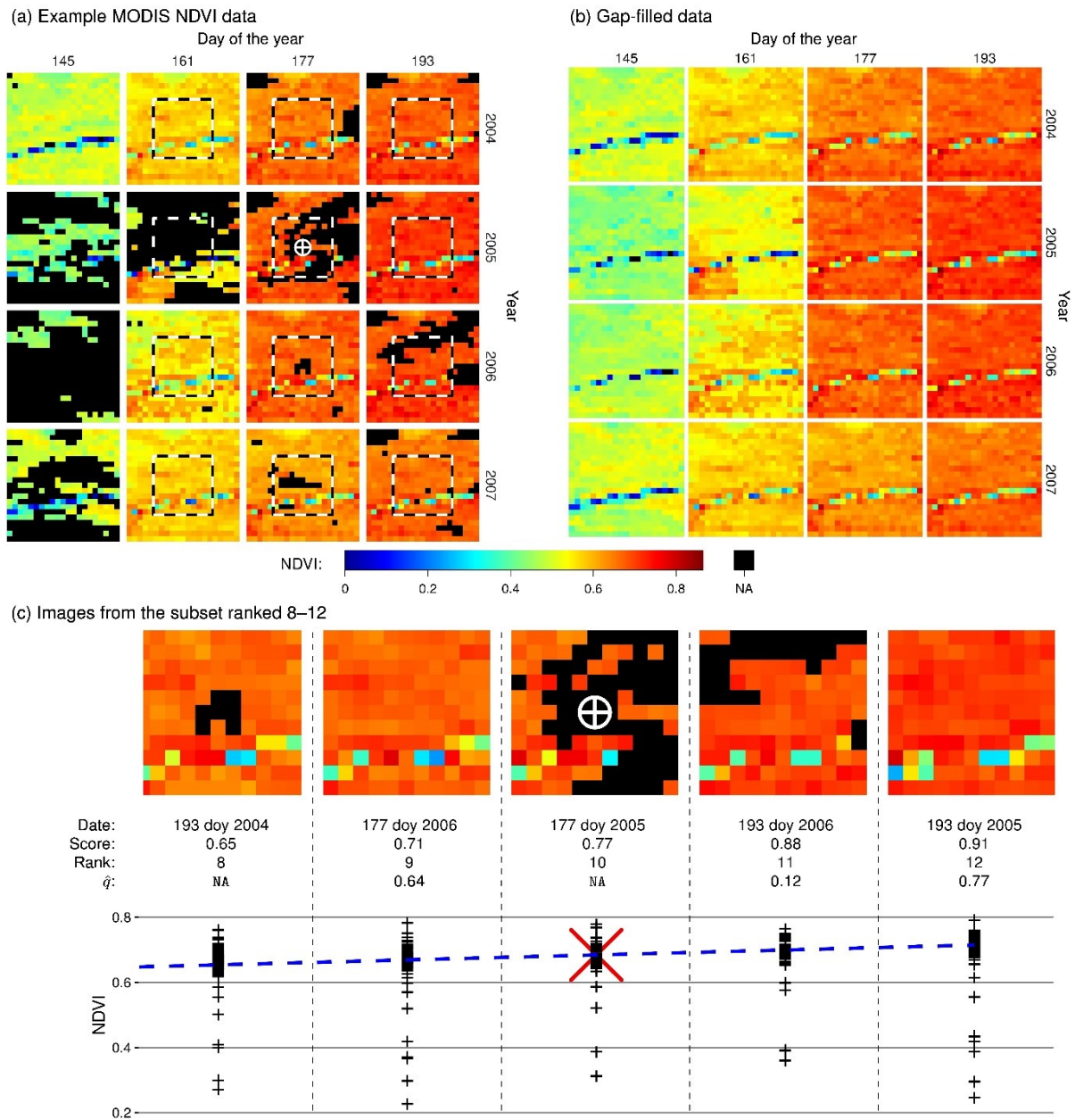

*Figure 1: Panel (a): example MODIS NDVI data. The depicted images have a spatial extent of $21 \times 21$ pixels and they were observed at $16$ points in time at $4$ different days of the year. The $1603$ missing values are depicted as black pixels. To predict the pixels of the $177th$ day of the year $2005$ marked with a white cross, the subset marked with dashed squares is considered. Panel (b): gap-filled data. Panel (c): insights in the fitting procedure of the value from panel (a) marked with a white cross are given. Depicted are the subsets marked with the dashed lines from Panel (a) ranked $8$–$12$. While the top row depicts the*

*images of that subset, the bottom row shows a scatted plot of the corresponding NDVI values. The dashed blue line is the quantile-regression fit and the red cross is the predicted NDVI value.*

In general, the target satellite products comprise large amounts of data. Therefore, a gap-fill procedure of practical significance has to be efficient in terms of computation resources and scalable in the sense that the gap-fill task can be distributed to several computing units. In order to achieve this, the presented gap-fill procedure predicts each missing value separately by taking only a subset of the data into account. The prediction has two main steps: (1) subset the data and (2) predict the missing value based on that subset. With that, the algorithm is adaptive and can recover local features of the data. This is useful if, for example, a region of the data exhibits a temporally shifted seasonal pattern or a long-term temporal trend. Note that many different algorithms having the subset-predict structure can be constructed.

In the following, the gap-fill procedure is described for one missing value. To fill an entire dataset, the procedure is repeated for all missing values. The gap-filled version of the example dataset is shown in panel (b) of Figure 1. In the first step, the data are subset to a neighborhood around the missing value containing the relevant information for the prediction. This subset is defined as four dimensional array including $\lambda_1 = \lambda_2 = 5$ pixels in all spatial directions from the missing value and images being not further apart than $\lambda_3 = 1$ step in both direction of the seasonal index and $\lambda_4 = 5$ years in both direction of the year index. Note that $\lambda_1, \dots, \lambda_4$ and the later introduced $\theta_1, \theta_2, \theta_3$ are tuning parameters and may be altered in order to optimize the procedure for a given dataset. The choice of these parameters can be justified via cross-validation. In Panel (a) of Figure 1, the missing pixel of interest is marked with a white cross and the corresponding subset is depicted with dashed squares. The following two criteria are verified to assess whether the subset contains enough non-missing values: (C1) the subset must contain at least $\theta_1 = 5$ non-empty images, and (C2) the image in the subset containing the missing pixel must have at least $\theta_2 = 25$ non-missing values. If one of those criteria is not met, the spatial extent of the subset is increased (by increasing the values of $\lambda_1$ and $\lambda_2$) until both criteria are fulfilled. With that mechanism, the spatial extent of the subset is adapted to the local distribution of missing values.

In the second step, the missing value is predicted based on the described subset. The latter is interpreted as a collection of images, and we rank them according to the intensities of their values. The rationale behind this is that the ranked series of images imitates the seasonal evolution of the spatial field with an artificially high temporal resolution. The ranking is based on an algorithm that scores the images via pixel-wise comparisons of non-missing values. The main assumption of the scoring is that the values of the images in the subset have a similar but potentially shifted spatial distribution. This is exploited to construct an algorithm, which is only marginally affected by missing values. In the data example, a subset of the ordered images and their scores are depicted in Panel (c) of Figure 1. There, the missing pixel of interest is marked with a white cross and contained in the center image ranked 8th. The bottom row of the same panel displays the ordered images as a scatter plot, having the estimated rank on the $x$-axis and the observed NDVI values on the $y$-axis.

To finally obtain the prediction of the missing value, linear quantile-regression is used. The regression has the NDVI values as a response and an intercept as well as the ranks of the image as linear predictors. With that, the ranked images serve as a non-parametric seasonality adjustment. The quantile of interest is estimated from the images that have an observed value at the spatial location of the missing value. We require at least $\theta_3 = 2$ such values to estimate the quantile. If the criterion is not met, spatially neighboring pixels are considered too. The quantile is estimated by evaluating the empirical cumulative distribution function at each such value and averaging these results. In panel (c) of Figure 1 the quantile of the missing pixel was estimated to be the $47\%$-quantile. The corresponding regression line is depicted as dashed blue line and the predicted NDVI value is a red cross.

The described gap-filling approach is implemented in the programming languages R/C++ and is available as open-source R package at http://cran.r-project.org/package=gapfill (Gerber, 2016; R Core Team, 2016). The program features a flexible design allowing the user to optimize the gap-fill procedure for specific datasets and to construct new gap-fill algorithms based on the subset-predict framework with little effort. Gap-filling can be executed in parallel via interfaces to openMP and MPI back-ends making the software suitable for large datasets. More information about the usage and implementation of the R package is given in the Section S2 of the supplementary material and in the reference manual of the R package.

## 2.2. Formal description

This section provides a technical description of the gap-fill procedure and can be skipped without loss of the general idea. Table 1 gives an overview of the used mathematical symbols. Let $z = z[x, y, s, a] \in \mathbb{R} \cup \{NA\}$ denote a value of the data that is either observed or missing (NA). The indices $x \in \{1, \ldots, N_x\} = I_x$ and $y \in \{1, \ldots, N_y\} = I_y$ describe the spatial location of $z$, and $s \in \{1, \ldots, N_s\} = I_s$ is a seasonal index describing the temporal position of $z$ within the year $a \in \{1, \ldots, N_a\} = I_a$. Moreover, let $\mathcal{A} = \{z : x \in I_x, y \in I_y, s \in I_s, a \in I_a\}$ denote the four dimensional input data array of interest having $N = N_x \times N_y \times N_s \times N_a$ elements in total. In the example introduced in Section 2.1 and Figure 1, the spatial extent of $\mathcal{A}$ is $N_x = N_y = 21$ and the temporal extent is $N_s = N_a = 4$.

Table 1: List of mathematical objects used to describe the algorithm.

| Notation | Explanation |
| --- | --- |
| $z = z[x, y, s, a] \in \mathbb{R} \cup \{NA\}$ | One value of the data (observed or missing) |
| $x \in \{1, \ldots, N_x\} = I_x$ | $x$ coordinate of the spatial position of $z$ |
| $y \in \{1, \ldots, N_y\} = I_y$ | $y$ coordinate of the spatial position of $z$ |
| $s \in \{1, \ldots, N_s\} = I_s$ | Temporal position of $z$ within a year |
| $a \in \{1, \ldots, N_a\} = I_a$ | Temporal position of $z$ indicating the year |
| $\mathcal{A} = z[\cdot, \cdot, \cdot, \cdot]$ | Input data containing observed and missing values |
| $z_0 = z_0[x_0, y_0, s_0, a_0] \in \mathcal{A}$ | One missing value of the data |
| $B(z_0) \subseteq \mathcal{A}$ | Neighborhood around $z_0$ |
| $z' = z'[x', y', r'] \in B'$ | Element in the projected neighborhood $B'$ |
| $f : B \to \mathbb{R} \cup \{NA\}, f(B) = \hat{z}_0$ | Function returning the predict value or NA |
| $\lambda_1, \lambda_2, \lambda_3, \lambda_4, \theta_1, \theta_2, \theta_3 \in \mathbb{N}$ | Tuning parameters of the gap-fill procedure |

To predict the missing values in $\mathcal{A}$, the following subset-predict procedure is repeated for all missing values. Let $z_0 = z_0[x_0, y_0, s_0, a_0] \in \mathcal{A}$ denote one such missing value. In the subset step, a suitable neighborhood $B = B(z_0) \subseteq \mathcal{A}$ around $z_0$ is selected. In the prediction step, the missing values is predicted based on $B$. We consider a neighborhood as suitable, if it contains enough non-missing values to predict the value of $z_0$. Since the missing values are often unevenly distributed in $\mathcal{A}$, $B$ may be different depending on the position of $z_0$. To find a suitable $B$, an iterative procedure is employed, which is formalized as follows. Let $B = B(z_0, i)$ depend on the parameter $i \in \mathbb{N} \cup \{0\}$ so that for increasing $i$ s the neighborhood increases or at least changes. Furthermore, define the prediction function $f: B \to \mathbb{R} \cup \text{NA}$ such that it returns the predicted value $f(B) = \hat{z}_0$, if $B$ is suitable and $f(B) = \text{NA}$ otherwise. The idea is to start with $i = 0$ and to try different neighborhoods $B$ by increasing $i$ until $f$ returns the predicted value $\hat{z}_0$.

### 2.2.1. Subset to a neighborhood

To be more specific about the search strategy for suitable subsets, let $B(z_0, i, \lambda_1, \lambda_2, \lambda_3, \lambda_4)$ be a four dimensional box around the missing value. The following definition shows that the spatial extent increases as $i$ increases.

$$B(z_0, i, \lambda_1, \lambda_2, \lambda_3, \lambda_4) = \{z[x,y,s,a] \in \mathcal{A} : x_0 - (\lambda_1 + i) \leq x \leq x_0 + (\lambda_1 + i), y_0 - (\lambda_2 + i) \leq y \leq y_0 + (\lambda_2 + i), s_0 - \lambda_3 \leq s \leq s_0 + \lambda_3, a_0 - \lambda_4 \leq a \leq a_0 + \lambda_4\} \quad (1)$$

The parameters $\lambda_1, \dots, \lambda_4$ define the initial and minimal extent of $B$. In case where $z_0$ is not close to a boundary of $\mathcal{A}$ the spatial extent of $B$ is $(2\lambda_1 + i) \times (2\lambda_2 + i)$ pixels.

### 2.2.2. Predict based on a subset

The first task of the prediction function $f$ is to decide whether to return NA or the predicted value $\hat{z}_0$. In order to return $\hat{z}_0$, both of the following criteria need to be fulfilled:

(C1) $B$ must contain at least $\theta_1 \in \mathbb{N}$ non-empty images.

(C2) The image in $B$ containing the missing pixel must have at least $\theta_2 \in \mathbb{N}$ non-missing values.

For $\theta_1 = 5$ and $\theta_2 = 25$ the subset depicted in Panel (a) of Figure 1 fulfills both the criteria (C1) and (C2). To explain the actual prediction of a missing value $z_0$, let $B$ be a neighborhood of $z_0$, which fulfills the criteria (C1) and (C2). By construction $B$ is a four dimensional array, which we map to a three dimensional array $B'$ by seeing it as a collection of $N'_r$ non-empty images with no particular temporal ordering. Hence, each of the $N'_x \times N'_y \times N'_r$ elements $z'[x', y', r'] \in B'$ can be addressed with three indices, namely $x' \in \{1, \ldots, N'_x\} = I'_x$, $y' \in \{1, \ldots, N'_y\} = I'_y$ and $r' \in \{1, \ldots, N'_r\} = I'_r$. Next, the $N'_r$ images of $B'$ are scored according to the values of their non-missing pixels. We assume that the images in $B'$ have a similar but potentially shifted distribution of values and exploit this to construct an algorithm, which is only marginally affected by the occurrence of missing values. The algorithm is based on pixel-wise comparisons of each of the $N'_r$ images in $B'$ with all $N'_r - 1$ other images in $B'$. Algorithm 1 describes how the score of one image $k \in I'_r$ is obtained. Note that in the following, mean($X$) is the function that returns the mean of all non-missing values of $X$. By repeating the procedure $N_r$ times, all images are scored and subsequently ranked according to their scores. The ranked images of the example dataset, as well as their scores, are shown in Panel (c) of Figure *1*.

*Algorithm 1: Score the k-th image in $B'$ relative to the other images in $B'$.*

| | | |
|---|---|---|
| 1: | Define $M$ as a $N'_x \times N'_y$ matrix, initialized with NA. | |
| 2: | Define $V$ as a vector of length $N'_r$, initialized with NA. | |
| 3: | for $r' \in I'_r \setminus \{k\}$ do | |
| 4: |   for $x' \in I'_x, y' \in I'_y$ do | |
| 5: |     if $z'[x', y', k]$ and $z'[x', y', r']$ are not NA **then** | |
| 6: |       if $z'[x', y', k] > z'[x', y', r']$ **then** | |
| 7: |         $M[x', y'] \leftarrow 1$ | |
| 8: |       else | |
| 9: |         $M[x', y'] \leftarrow 0$ | |
| 10: |       end if | |

| 11: |     end if |
|---|---|
| 12: | end for |
| 13: | $V[r'] \leftarrow \text{mean}(M)$ |
| 14: | end for |
| 15: | **Return** mean($V$) |

We use the dot notation to indicate collections of points; for example, the entire neighborhood $B'$ is denoted with $z'[\cdot,\cdot,\cdot]$ and the image in $B'$ with rank $r'$ is denoted with $z'[\cdot,\cdot,r']$. The prediction of $z_0 = z'_0$ is obtained by fitting a quantile regression to the values of $B'$ having an intercept and the rank $r'$ of the images as linear predictors. The following equation describes the model for the $\alpha$-quantile $Q(\alpha)$.

$$Q(\alpha \mid r') = \beta_0(\alpha) + r'\beta_1(\alpha)$$

Let $\hat{\alpha}_0$ be an estimate of the quantile of the missing value $z'_0$ relative to $z'[\cdot,\cdot,r'_0]$, i.e., the image of $B'$ containing the missing value. Given $\hat{\alpha}_0$ is known, we can estimate the coefficients $\hat{\beta}_0(\hat{\alpha}_0)$ and $\hat{\beta}_1(\hat{\alpha}_0)$ by solving a minimization problem (Koenker, 2005; McMillen, 2012). The predicted value of $z'_0$ is then

$$\hat{z}'_0 = \hat{\beta}_0(\hat{\alpha}_0) + r'_0 \hat{\beta}_1(\hat{\alpha}_0).$$

To estimate $\hat{\alpha}_0$, let $\hat{F}_{r'}(x)$ denote the empirical cumulative distribution function estimated from the non-missing values of $z'[\cdot,\cdot,r']$. Moreover, let $\hat{\alpha}_{r'} = \hat{F}_{r'}(z[x_0',y_0',r'])$ be the estimated quantile of the value $z[x_0',y_0',r']$ relative to the image $r'$. We estimate the quantile of interest $\alpha_0$ as the mean of all defined values of $\{\hat{\alpha}_{r'}: r' \in I'_r\}$. Note that some of those values may be undefined because of missing $z'[x'_0,y'_0,r']$ values. We require at least $\theta_3$ defined values in $\{\hat{\alpha}_{r'}: r' \in I'_r\}$. If this criterion is not met, all $\hat{F}_{r'}$, $r' \in I'_r$ are also evaluated in a spatial neighborhood of $(x'_0, y'_0)$. A detailed description of the estimation of $\alpha_0$ is given in Algorithm 2.

*Algorithm 2: Estimate the $\alpha_0$-quantile of the missing value relative to the image $z[\,\cdot\,,\,\cdot\,,r_0]$.*

| | |
|---|---|
| 1: | Define $\boldsymbol{V}$ as a vector of length $N'_r$, initialized with NA |
| 2: | Set $i \leftarrow 0$, $D \leftarrow z'[x'_0, y'_0, \cdot\,]$ |
| 3: | **while** the number of non-missing values in $D$ is less than $\theta_3$ **do** |
| 4: | $\quad i \leftarrow i + 1$ |
| 5: | $\quad D \leftarrow \{z'[x',y',\cdot\,] \in B' : x'_0 - i \leq x' \leq x'_0 + i, y'_0 - i \leq y' \leq y'_0 + i\}$ |
| 6: | **end while** |
| 7: | Define $A$ as an array with extent $(2i+1) \times (2i+1) \times N'_r$, initialized with NA |
| 8: | **for** $r' \in I'_r$ **do** |
| 9: | $\quad$ Estimate $\widehat{F_{r'}}(\cdot)$ from $z'[\,\cdot\,,\,\cdot\,,r']$ |
| 10: | $\quad$ **for** $x' \in \{-i, \ldots, i\}, y' \in \{-i, \ldots, i\}$ **do** |
| 11: | $\quad\quad$ **if** $z'[x'_0 + x', y'_0 + y', r']$ not is NA **then** |
| 12: | $\quad\quad\quad A[x'_0 + x', y'_0 + y', r'] \leftarrow \widehat{F_{r'}}(z'[x'_0 + x', y'_0 + y', r'])$ |
| 13: | $\quad\quad$ **end if** |
| 14: | $\quad$ **end for** |
| 15: | $\quad V[r'] \leftarrow \text{mean}(A[\,\cdot\,,\,\cdot\,,r'])$ |
| 16: | **end for** |
| 17: | **Return** $\text{mean}(V)$ |

## 2.3. Prediction uncertainties

Uncertainty estimates of the predicted values are essential when using them to derive conclusions in further analyses. Statistical theory provides ways to quantify uncertainty through the estimation of prediction variability and confidence intervals. Possible strategies that can be applied to the proposed gap-fill method are based on resampling techniques and cross-validation. While the former is computationally expensive, and therefore difficult to apply to a large dataset, the latter was applied in a similar setting (Weiss et al., 2014). However, both approaches are inaccurate, if the underlying assumptions about the data are not met. For example, the commonly made "missing at random" assumption is not met in the data considered in Section 3.1, because low NDVI values have a larger probability to be missing.

Besides that, it is interesting to study the magnitude of the uncertainties introduced by the different steps of the procedure; the latter are (1) choosing the size of the initial subset around the missing value, (2) ranking the images of the subset based on Algorithm 1, (3) estimating the quantile of the missing values based on Algorithm 2, and (4) estimating the parameters of the quantile regression. We assessed the uncertainties introduced in step (1) by running the gap-fill procedure with all possible initial sizes of the spatial subset. Then a $90\%$ prediction interval for each missing value is obtained by considering the $5\%$ and the $95\%$ quantile of the predicted values. The uncertainty of step (2) is calculated via permutations of the ranked images. More precisely, we approximate all possible permutations of the ranks by changing the rank of the image containing the missing pixel to all possible positions. The motivation for this approximation is that the permutation of the image containing the missing value has a much larger influence on the predicted value compared to permutations among other images. The predicted values for all such permutations are then calculated and a $90\%$ prediction interval is constructed by considering the $5\%$ and the $95\%$ quantile thereof. For step (3), we derive a $90\%$ prediction interval by considering the $5\%$ and the $95\%$ quantiles of the estimated quantiles in $V$ of Algorithm 2. Finally, the uncertainty introduced in step (4) is assessed by calculating a $90\%$ prediction interval based on bootstrap methods.

Estimates of the prediction uncertainties of the gap-filled values could be derived by combining the uncertainties of all four previously described steps. However, doing so in a meaningful way is not straightforward due to possible interactions and elimination effects among them. Nevertheless, we combine the uncertainties from step (2) and (3) in one prediction interval. This prediction interval reflects the local spatial and temporal heterogeneity of the data around the predicted value. An evaluation of the properties of that interval and its practical relevance is given in Section 3.3.2 and 4.2.

# 3. Data and validation method

Several test scenarios were constructed and the predicted values, together with their uncertainty components, were investigated. In addition, the accuracy of the gap-filled values was compared against those of two alternative gap-fill procedures.

## 3.1. Data

We considered the MODIS satellite product (MOD13A1), which is part of the MODIS vegetation index product (MOD13) (Justice et al., 2002). It is a land surface product based on pre-composited 8-day MODIS Level-2G surface reflectance data, which have been further composited to obtain the final resolution of circa 16 days and 500 m (Didan et al., 2015; van Leeuwen et al., 1999). The NDVI layer can be used to describe vegetation activity (Huete, et al., 2002). We used the quality assignments pixel reliability layer to subset the NDVI data to values classified as "good data" (Roy et al., 2002).

To illustrate the gap-fill procedure, the NDVI data from the years 2004 to 2009 were considered as the subset of the region of northern Alaska as shown in Figure 2. Due to the high latitudes ranging from $66°$ north to more than $71°$ north, the NDVI values exhibit a strong seasonal component. That is reflected in both the NDVI values and in the number of available values classified as "good data". Especially during wintertime, little data are available because of missing sunlight and snow cover. Therefore, we restrict the analysis to the seasonal period starting on the 145th day of the year (about Mai 24) and ending on the 257th day of the year (about September 13) featuring 8 images per season. With that, the data of each considered day of the year have at least $30\%$ of the values classified as "good data". The MOD13A1 data were downloaded in $6$ spatial tiles and merged to one single image per considered day of the year using the R package MODIS (Mattiuzzi, 2015), which interfaces the MODIS reprojection tool (Dwyer & Schmidt, 2006). Furthermore, the data were transformed from the sinusoidal to the geographic map projection (WGS84). The R-packages *raster* (Hijmans, 2015), *sp* (Bivand, Pebesma, & Gomez-Rubio, 2013; Pebesma & Bivand, 2005), *fields* (Nychka, Furrer, & Sain, 2015), *lattice* (Sarkar, 2008), and *ggplot2* (Wickham, 2009) were used to handle and visualize the data.

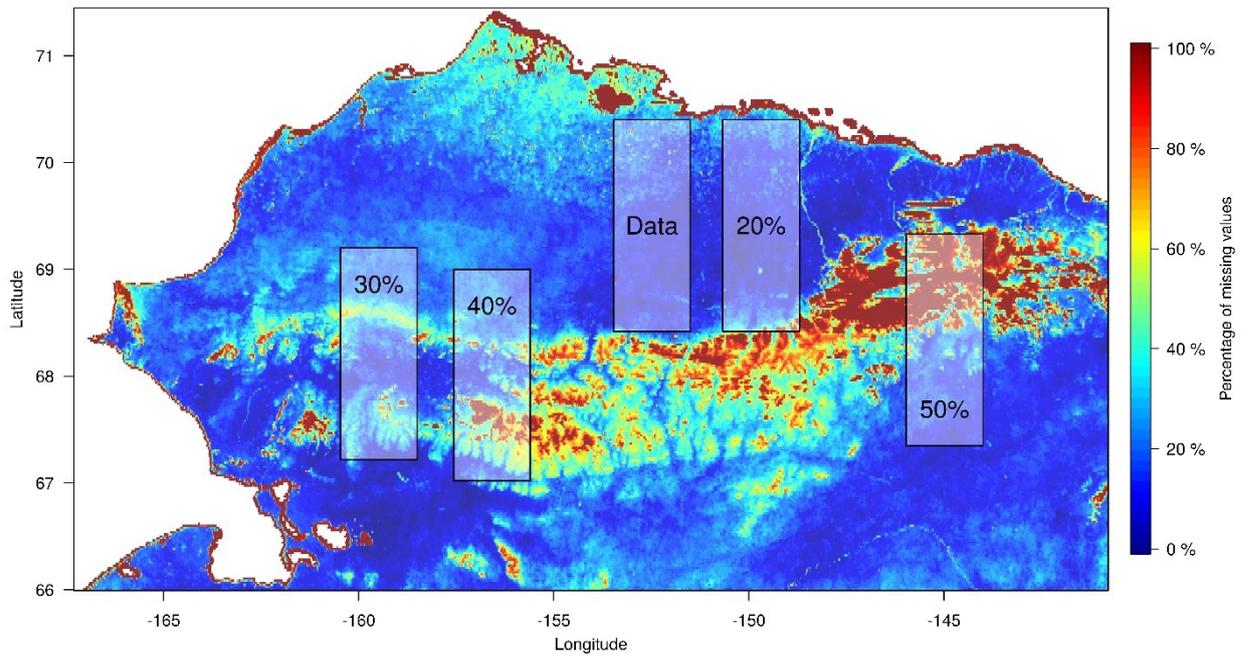

*Figure 2: A map of the study region of northern Alaska. The colors indicate the percentage of missing pixels in all 48 considered days of the year. To construct validation scenarios, we consider the NDVI values of the $100 \times 100$-pixels region labeled with "Data". That subset exhibits relatively few (about 12%) missing values. Test scenarios with 20%, 30%, 40%, and 50% missing pixels were obtained by artificially removing values from the "Data" region according to the patterns of missing values observed at the $100 \times 100$-pixels regions labeled with "20%", "30%", "40%", and "50%" respectively. Note that the $100 \times 100$-pixels regions are depicted as rectangles, as opposed to squares, because of the chosen geographic map projection.*

### 3.2. Test scenarios

To study the performance of the gap-fill software, we constructed four tests scenarios based on the MOD13A1 data. Using real data, as opposed to simulated data, has the advantage that the scenarios come close to the use-case of interest. The scenarios were built by extracting a $100 \times 100$-pixel subset of the data described in Section 3.1, while keeping the temporal structure of 6 years with 8 images per year unchanged. The geographical location of the subset is depicted in Figure 2 as the rectangle labeled with "Data". The rectangular area was selected to ensure that the resulting dataset has relatively few missing values (about 12%) and the values reflect typical features of NDVI datasets in high latitudes. Two of these features are the latitudinal gradient manifesting itself though lower NDVI values in the northern regions and low NDVI values that are caused by surface water. By artificially removing values from that subset (setting them to NA), validation scenarios exhibiting 20%, 30%, 40%, and 50% missing values were retrieved. In this way, we know most of the actually observed (true) values of the test scenario and can compare the values from the gap-filled versions against them. The removal of NDVI values was performed according to patterns of missing values observed at other locations of the Alaska dataset. They are depicted in Figure 2 as the rectangles denoted with "20%", "30%", "40%", and "50%". As such, the patterns of missing values in the test scenarios have a realistic spatio-temporal structure. The NDVI values of the four test scenarios and some summary figures thereof are depicted in Figures S1–S5 of the supplementary material.

In addition, a scenario consisting of the entire spatial extent of northern Alaska, as shown in Figure 2, was compiled. While the temporal dimensions of that scenario remained unchanged, the size of the images is increased to 271'819 pixels. 3'696'691 (28%) of the values in that scenario are missing. The scenario is shown in Figure S10 of the supplementary material.

### 3.3. Evaluation of the gap-fill method

#### 3.3.1. Predictions accuracy

A first evaluation criterion for gap-fill methods is the proportion of values that remained missing after applying them. Good gap-fill algorithms are capable of predicting many missing values of a dataset. A second criterion is the visual examination of the filled values, which helps to detect artificial patterns introduced by the prediction procedure. More objective measurements of the prediction accuracy can be made when considering the fill values of the test scenarios described in Section 3.2. Here most of the true values are known and the deviation of the fill values from them can be quantified. We summarized the prediction accuracy with the root mean squared error (RMSE), defined as $\sqrt{\sum_{i=1}^{n}(\hat{z}_i - z_i)^2/n}$, where $n$ is the number of (non-missing) predictions and $z_i$ and $\hat{z}_i$ denote the true and the predicted values, respectively.

### 3.3.2. Uncertainty assessment

The prediction algorithm exhibits four main steps that are linked to the uncertainties of the predicted values (Section 2.3). To compare the magnitudes of these individual uncertainty contributions, the lengths of the corresponding prediction intervals are summarized. Due to the considerable computational workload of that task, this investigation was performed on the images of the days $145$ and $161$ of the year from the test scenario with $40\%$ missing values. The properties of the 90% prediction interval combining the uncertainties from step (2) and (3) were assessed by investigating the spatial and temporal distribution of the prediction interval lengths and by calculating the coverage rate of the intervals, i.e., the proportion of intervals that cover the observed value (assumed to be true). This part of the uncertainty assessment was performed using the entire test scenario with $40\%$ missing values.

### 3.3.3. Comparison with TIMESAT and Gapfill-Python

We compare the results from the described gap-fill procedure – denoted as the corresponding R package with "*gapfill*" – against two alternative procedures, namely the gap-fill algorithm presented in Weiss et al. (2014) and the temporal interpolation methods implemented in the software TIMESAT (Eklundh & Jönsson, 2015; Jönsson & Eklundh, 2004). The former belongs to the class of algorithms that reconstruct a complete data product from the observed values (Section 1.2.2). It applies one of two different prediction algorithms depending on the amount of missing values and exploits both the temporal and the spatial structure of the data. A python notebook is available at github.com/malaria-atlas-project/modis-gapfilling and provides an implementation of the procedures in Python (Foundation, 2015) and C. The code was downloaded on August 15, 2015. In the following, we will refer to that software as "Gapfill-Python". While that method was published in 2014, the TIMESAT (version 3.2) software is more than ten years older and well established. It is implemented in Fortran and comes with a MATLAB (MATLAB, 2014) interface featuring a graphical user interface. The software and the documentation thereof is available at web.nateko.lu.se/timesat/. The main purpose of the software is to analyze time series of satellite data by extracting seasonal parameters from a smoothed version of the time series. All calculations treat the pixel-wise time series separately, and hence, do not exploit the spatial dependency in the data. The smoothing part of the method makes the analysis, to some extent, robust to outliers and missing values. The method therefore belongs to the class of methods that handle missing values through robust analysis techniques (Section 1.2.1). Although the software was designed to use the smoothed time series in conjunction with the extraction of phenomenological parameters, the smoothing part of the algorithm can be used for gap-filling and the gap-filled time series were used to assess the performance of different types of smoothers (Atkinson et al., 2012; Hird & McDermid, 2009; Verger et al., 2013).

All considered gap-fill procedures have several tuning parameters, which influence the prediction process and the accuracy of the predictions. Although, we tried to find good parameter configurations for the presented software, it may be that the results improve with other settings. Nevertheless, the presented comparisons give an impression of the performance. The tuning parameters for the gap-fill procedure are the same as in Section 2 except of the parameters $\lambda_1$ and $\lambda_2$ which we set to 10. The R-code to run the gap-fill algorithm is given in Listing S1 of the supplementary material. Gapfill-Python has about 16 parameters to be set. The most important ones are those controlling the search of informative pixels and the "de-speckle" procedure (see the Python-notebook for more information). The used parameters are given in Listing S2 of the supplementary material. The parameters of TIMESAT are described in its software manual (Eklundh & Jönsson, 2015). We chose to fit a "double logistic" smoothing function, which is recommended for NDVI values in high latitudes with many missing values (Beck, Atzberger, Høgda, Johansen, & Skidmore, 2006). Listing S3 of the supplementary material shows the complete configuration file.

# 4. Results

## 4.1. Predictions accuracy

We applied the proposed gap-fill procedure to the test scenarios described in Section 3.2. The resulting gap-filled images are shown in the Figures S6–S9 of the supplementary material and the left panel of Figure 3, which depict the predicted values of day 177 of the year 2006 for the test scenario with 40% missing values. The gap-fill procedure did return predicted values for all missing values in all test scenarios. A visual examination of the gap-filled images did not reveal any artificially introduced spatial pattern. Moreover, the images reconstruct the spatial distributions of the NDVI values well; this includes small-scale features such as, e.g., the band of low NDVI values crossing the image from the west to the east, which is present in all images. The seasonal variation of the NDVI values and deviations of some images thereof (see e.g., the image of the 241 day of the year 2009) are recovered in the predicted images.

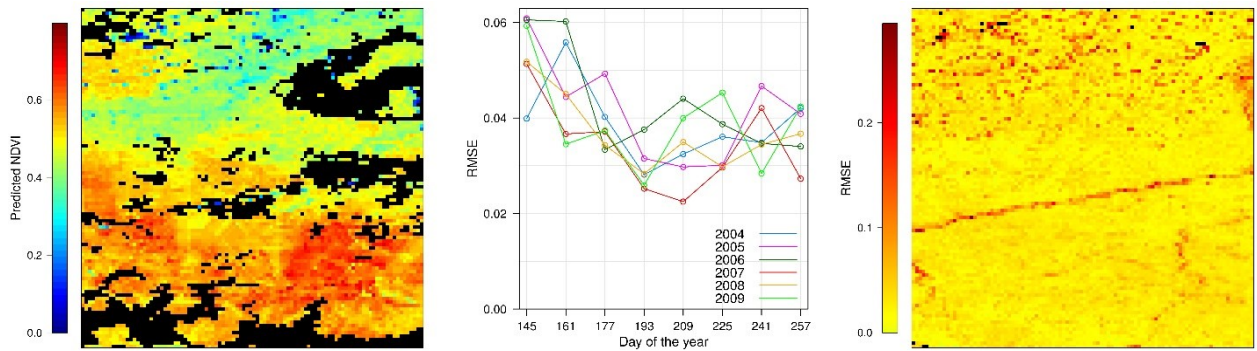

*Figure 3: Predictions and accuracy measurements for the scenario with 40% missing values. Left panel: predicted NDVI values for the day 177 of the year 2006. The observed values from that image are shown as black pixels. Middle panel: RMSE for the indicated dates. Right panel: spatial distribution of the RMSE. Missing RMSE values caused by the design of the test scenario are depicted as black pixels.*

To assess the temporal variation of the prediction accuracy, the average RMSE for each image of the scenario with 40% missing is shown in the middle panel of Figure 3. It can be seen that the RMSE is larger for early days of the year. This is in accordance with the observations that images at the beginning of the season tend to have more missing pixels and their values exhibit larger variability in time compared to other images (see also Figure S1 in the supplementary material). The right panel of Figure 3 depicts the spatial distribution of the pixel-wise average RMSE, which resembles the spatial distribution of the temporal variation in the data as depicted in the second panel of Figure S1 of the supplementary material. This is expected, because pixels with a large variability in time are more difficult to predict.

Another way to study the prediction accuracy is to plot the true values against the predicted values as shown in the upper panels of Figure 4. Most of the true values are between $0.3$ and $0.8$. In that region, the predicted values are scattered around the red line indicating that they are near the true ones on average. Pixels with values below $0.5$ have lower prediction accuracy. This is in accordance with the observation that those pixels tend to have large variance and are therefore naturally more difficult to predict. As expected, the deviation of the predicted values from the true values increases with larger percentages of missing pixels. This can also be seen in the bottom panels of Figure 4, where the histograms of the prediction errors (predicted minus true values) show a wider distribution with increasing percentages of the missing pixels. While the median differences are located at zero, the distributions of the differences are slightly positively skewed. This might be caused by the negative skewed distribution of the true values and their restriction to the interval $[0,1]$.

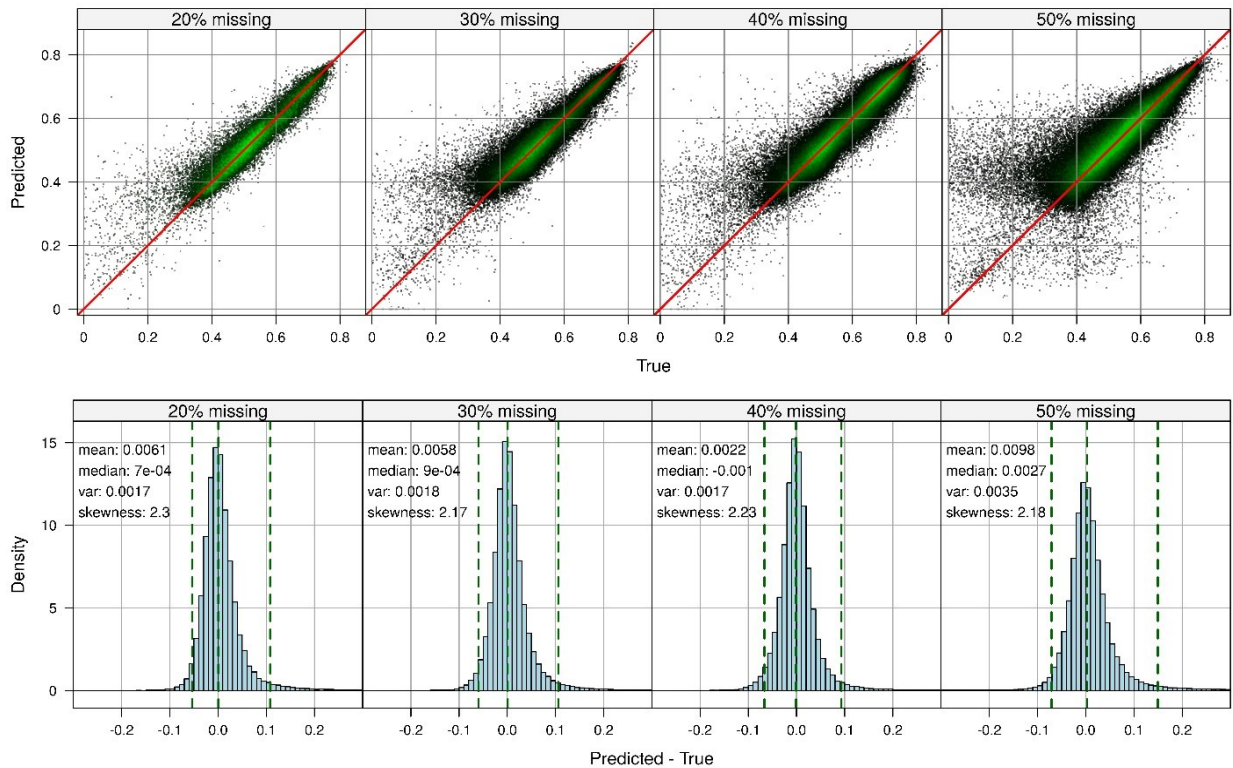

*Figure 4: Accuracy of the gapfill predictions for the four test scenarios of the validation study. Upper panels: scatterplots of the predicted values ($y$-axis) versus the true values ($x$-axis). The green color shading indicates regions with a large density of points (light green corresponds to $100$ overlaying points). Bottom panels: histograms of the differences between the predicted and the true values. The dashed green lines indicate the $2.5\%$, $50\%$, and $97.5\%$ quantiles.*

In addition, *gapfill* was applied to the entire spatial region of northern Alaska depicted in Figure 2. The images of the gap-filled data are shown in Figures S11 of the supplementary material. Again, all missing values were filled and a visual inspection of the data did not reveal any artificially introduced patterns.

### 4.2. Uncertainty assessment

The widths of the prediction intervals, corresponding to the four main steps of the prediction algorithm, summarize their uncertainty contribution. The left panel of Figure 5 depicts summary statistics of these widths as boxplots revealing that the sorting step (2) (**Algorithm 1**) introduced the larges uncertainties, followed by the estimation of the quantile of step (3) (**Algorithm 2**). To investigate the properties of the prediction interval combining the uncertainties from step (2) and (3), the spatial distribution of the average prediction interval widths is shown in the middle panel of Figure 5. It exhibits similar spatial patterns as the standard deviation of the data (middle panel of Figure S1 of the supplementary material) and the spatial distribution of the average RMSEs (right panel of Figure 3). Since the seasonal variability of the prediction interval widths is larger, compared to the inter-annual variability, we only show the former in the right bottom panel of Figure 5. It has a U-shape, which is also observed in the distributions of the missing values (Figure S1 of the supplementary material) with some deviations thereof at the boundaries, i.e., the values of the 145th and the 257th day of the year. These deviations might be caused by the fact that we only consider a part of the seasonal cycle, and hence, have less information at the boundaries thereof. The overall coverage rate of the prediction interval for that scenario is 93%. This is, the prediction uncertainty is slightly overestimated on average. The average coverage rate per day of the year is depicted in the right panels of Figure 5.

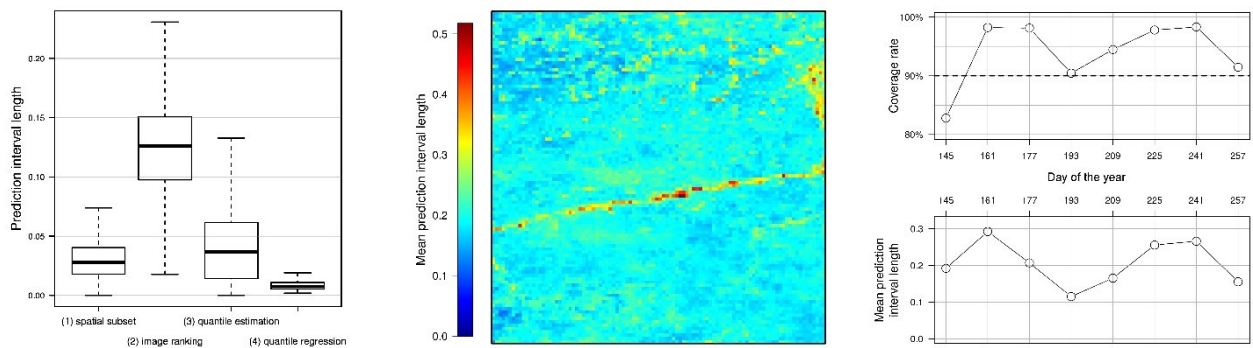

*Figure 5: Left panel: uncertainty contribution from the indicated four steps of the gap-fill procedure. Middle panel: spatial distribution of the average width of the 90% prediction intervals for the test scenario with 40% missing values. The intervals are based on the uncertainties from **Algorithm 1** and **Algorithm 2**. Right panel: the corresponding average interval widths and coverage rate per day of the year.*

### 4.3. Comparison with TIMESAT and Gapfill-Python

When comparing the predictions of *gapfill* with those from Gapfill-Python and TIMESAT, one performance measurement of interest is the ability to fill gaps in scenarios exhibiting many missing values. To investigate that, the number of predicted values, and the percentage of predicted values relative to the total number of missing values, is shown in the columns "# filled" of Table 2. While *gapfill* predicted all missing values of the four test scenarios, Gapfill-Python and TIMESAT returned NA predictions for some values. The number of NAs in the predictions seems to increase with the number of NAs in the input data and are a considerable proportion (up to 94%) for the TIMESAT software.

*Table 2: The predicted values obtained with the gapfill, Gapfill-Python and TIMESAT are summarized in terms of the number of missing values in the prediction and RMSE $\times 10^3$. To get comparable results, the RMSE of the gapfill approach is also give for the subset of available predictions from Gapfill-Python (RMSE$_P$) and from TIMESAT (RMSE$_T$).*

|     | gapfill | | | | Gapfill-Python | | TIMESAT | |
| --- | --- | --- | --- | --- | --- | --- | --- | --- |
|     | #filled | RMSE | RMSE$_P$ | RMSE$_T$ | #filled | RMSE | #filled | RMSE |
| 20% | 92'822 (100%) | 41.80 | 42.06 | 41.10 | 90'307 (97%) | 45.00 | 59'948 (65%) | 83.43 |
| 30% | 147'827 (100%) | 42.54 | 42.39 | 37.09 | 146'686 (99%) | 45.54 | 42'892 (29%) | 71.43 |
| 40% | 192'456 (100%) | 41.34 | 40.98 | 36.41 | 169'998 (88%) | 42.49 | 31'279 (16%) | 71.93 |
| 50% | 240'326 (100%) | 59.58 | 44.94 | 37.24 | 134'540 (56%) | 45.61 | 14'127 (6%) | 86.09 |

The prediction accuracy in terms of the RMSE can only be calculated for non-missing predictions. Thus, the RMSE of the Gapfill-Python and the TIMESAT predictions are based on a subset of the values to predict only. To make the RMSE of those methods comparable to that of *gapfill*, we also calculated the RMSE of the *gapfill* method relative to the subsets of predicted values available for Gapfill-Python and TIMESAT. They are denoted with RMSE$_P$ and RMSE$_T$, respectively. The RMSEs given in Table 2 indicate that the *gapfill* predictions are the most accurate ones in all scenarios, although the predictions of the Gapfill-Python software have a comparable performance. In contrast, the RMSEs of TIMESAT are about two times higher compared to those of *gapfill*.

# 5. Discussion

The analysis of remotely sensed data with many missing observations is challenging. One way to handle missing values is to construct a complete dataset by predicting the missing values from the observed ones. Such an approach is presented in this study and implemented in the corresponding R package *gapfill*. The following four considerations influenced the development of the gap-fill method:

Firstly, to be of practical relevance the method has to be capable of handling large datasets, such as, e.g., MODIS NDVI products. This implies that classical geostatistical space-time models in the spirit of Stein (1999) would need sever modifications, since their computation workload typically exceeds the available resources by several orders of magnitude. One way to reduce the computational workload is to implement a purely algorithmic procedure as, e.g., presented by Weiss et al. (2014). While such approaches can achieve good performance in terms of prediction accuracy, uncertainty quantification is difficult. We therefore opted for a hybrid approach, which combines purely algorithmic elements (selection of a suitable subset; scoring of images) together with statistical methods (quantile regression; permutation tests). With that, the gap-fill method benefits from both aspects: The algorithmic components make it fast and scalable, whereas the statistics part provides tools to quantify uncertainties.

Secondly, an efficient software implementation of the gap-fill method is crucial to handle large datasets. Since nowadays many research institutes have access to powerful computers, this includes scalability of the algorithm so that the computational workload can be distributed among several computing units. The presented gap-fill algorithm achieves this with the subset-predict frame work, which handles each missing value separately and thereby enables parallelization. On the programming side, we employ the generic parallelization framework of the R package *foreach* (Analytics & Weston, 2015), which can be used to interface openMP and MPI parallel back-ends depending on the architecture of the available computer.

Thirdly, testing gap-fill algorithms with realistic scenarios was essential to develop an accurate and fast procedure. The chosen test scenarios feature actually observed MODIS NDVI data together with observed patterns of missing values and are therefore close to an actual use-case. In the development phase of the procedure, testing helped to detect alleged improvements, which turned out to be deteriorations with respect to computational speed or prediction accuracy. Also the comparison against established software provides valuable information for potential users. However, one needs to keep in mind that such comparisons are always relative to the choice of the test scenarios.

Fourthly, the gap-fill software should be flexible and user-friendly so that it can be tailored to specific features of different remote sensing data products. To that end, we provide the procedure as open-source R package *gapfill* (Gerber, 2016), which contains the R/C++ source code together with documentation and data examples. The structure of the package was kept as simply as possible to ease its usage. On the other hand, user can customize the essential parts of the algorithm by changing default parameters or by providing their own subset and/or predict functions; see Section S2 of the supplementary material for more information.

## 6. Conclusion

Since many analysis methods for remotely sensed data are designed for complete data, gap-filling missing values improves or enables data analysis in the presence of missing values. We presented such a gap-fill procedure and provide an open-source implementation thereof in the R package *gapfill*. While the software readily works for the presented data, its flexible design allows the users to tailor it to specific needs with little effort. With that it is a suitable tool to gap-fill many data products observed at regularly spaced points in time. Furthermore, the algorithm has statistical components, which enable uncertainty quantification. The performance of the predicted values was assessed using validation scenarios based on MODIS NDVI data featuring between 20% and 50% missing values. When compared with two alternative gap-fill methods, the presented method was able to handle data with a higher percentage of missing values and provided the most accurate prediction in terms of the RMSE.

# Acknowledgments

We thank Harry Gibson for sharing the Gapfill-Python software and for answering related questions. We acknowledge support of the University of Zurich Research Priority Program (URPP) on "Global Change and Biodiversity." Authors sequence is listed following the SDC approach (Tscharntke, Hochberg, Rand, Resh V. H., & Krauss, 2007).

# Acknowledgments